\documentclass[reprint,nofootinbib,showkeys,amsmath,amssymb,aps]{revtex4-1}

\usepackage{verbatim} 
\usepackage{color}
\usepackage{graphicx}

\usepackage{amssymb, latexsym}
\usepackage{amsmath} 
\usepackage{amsthm}
\usepackage{bm}
\usepackage[mathscr]{eucal}
\usepackage{enumerate}

\begin{document}

\begin{titlepage}

\begin{center}
\textbf{Ray-tracing semiclassical low frequency acoustic modeling with local and extended reaction boundaries}
\\Running tile: Ray-tracing semiclassical acoustic modeling
\vspace{10ex}

Rok Prislan \footnote{e-mail: rok@prislan.net}\\
Department of Physics, Faculty of Mathematics and Physics, University of Ljubljana, Jadranska 19, SI-1000 Ljubljana, Slovenia\\\vspace{5ex}

Daniel Sven\v{s}ek\\
Department of Physics, Faculty of Mathematics and Physics, University of Ljubljana, Jadranska 19, SI-1000 Ljubljana, Slovenia
\end{center}\vspace{3ex}

\end{titlepage}

\begin{abstract}

\noindent
The recently introduced acoustic ray-tracing semiclassical (RTS) method is validated for a set of practically relevant boundary conditions. 
RTS is a frequency domain geometrical method which directly reproduces the acoustic Green's function. As previously demonstrated for a rectangular room and weakly absorbing boundaries with a real and frequency-independent impedance, RTS is capable of modeling also the lowest modes of such a room, which makes it a useful method for low frequency sound field modeling in enclosures.
In practice, rooms are furnished with diverse types of materials and acoustic elements, resulting in a frequency-dependent, phase-modifying absorption/reflection. In a realistic setting, we test the RTS method with two additional boundary conditions: a local reaction boundary simulating a resonating membrane absorber and an extended reaction boundary representing a porous layer backed by a rigid boundary described within the Delany-Bazley-Miki model, as well as a combination thereof.
The RTS-modeled spatially dependent pressure response and octave band decay curves with the corresponding reverberation times are compared to those obtained by the finite element method. 
\end{abstract}

\maketitle

\setlength{\parindent}{5ex}

\section{Introduction}\label{sec:Introduction}

\noindent 
Room acoustics simulations have become a valuable design tool for acousticians to predict room acoustic parameters, model room modal shapes, optimize the positioning of public address systems etc. 
Notwithstanding their increasingly versatile applicability, these tools should be still used carefully. In fact, it is not in all circumstances that relevant results are obtained and caution as well as thorough understanding of the underlying principles of each modeling technique is required from the user. According to their fundamental physical assumptions, the simulation methods used in acoustics are primarily divided into \emph{wave-based} methods and \emph{geometrical} methods. 

Wave-based methods solve the complete wave equation for a given geometry and boundary conditions (BCs) and are therefore in principle able to model all properties of waves --- interference, diffraction, and scattering. In practice these effects are crucial at low frequencies, where boundary element methods (BEM) \cite{kawai:2007}, finite element methods (FEM) \cite{easwaran:1995}, and finite difference methods (FDM) \cite{kowalczyk:2008} are most commonly used. Unfortunately, at higher frequencies wave-based methods become over demanding in terms of processing time and memory usage \citep{desmet:2002,harari:1992} although efforts have been devoted to extend their computationally effective frequency range (e.g., \cite{mehra:2012}).

On the other side, geometrical methods assume that sound propagates from the source in straight lines, implying that diffraction/scattering effects are not included by definition. Generally, the geometrical modeling techniques are limited to large rooms/sufficiently high frequencies, where the sound field is not significantly shaped by individual room modes. A special, rather restricted but practically widely used subgroup of these methods are energy methods, which consider solely the propagation of sound energy implying that also interference effects are disregarded. These methods are used in commercial room acoustics simulation engines (e.g., ODEON \cite{Naylor}, CATT-Acoustic \cite{Dalenback}). The energy methods are complemented by the so-called phased geometrical methods in which phase information is included.

Various implementations of geometrical methods exist and their further differentiation is commonly performed on the basis of their stochastic nature (ray-tracing method \cite{krokstad:1968}), the presence of virtual sources (image source method \cite{allen:1979}) or other criteria. Each implementation has its advantages and disadvantages for specific types of problems; for recent reviews cf., e.g., refs.~\cite{vorlander:2013,savioja:2015,pompei:2015}. 

Generally, the inclusion of the phase information improves the quality of the results obtained at low frequencies. Suh and Nelson showed \cite{suh:1999} that the geometrically modeled impulse response agrees better with the experiment if the phase is included. Various aspects of phased image source implementations have been independently studied and compared with other modeling methods \cite{vorlander:1989,jeong:2008,dance:1995,aretz:2014,lam:2005}. We (Prislan \emph{et al.}) \cite{prislan:2016} have been concentrating on reproducing individually resolved resonances well below the \textit{Schr\" oder frequency} \cite{schroder:1954} by implementing the ray-tracing semiclassical (RTS) method which reproduces the Green's function, $G$, rigorously within the semiclassical (eikonal) approximation. Our study showed that the results obtained by RTS can correctly model the modal structure of a rectangular room including its lowest modes, which is decisive for using the method at low frequencies.

To achieve relevant results with room acoustics simulations it is also important to fully understand the implemented BC. In geometrical modeling the BC is described in terms of the boundary impedance which determines the reflection coefficient of the boundary as if it were an extensive surface. It is common to assume that the boundaries are \emph{locally reacting}, in which case the surface impedance is independent of the angle of incidence of the sound wave \cite{rossing:2007}. For locally reacting boundaries, the importance of the BC in a phased geometrical method was demonstrated by Jeong \cite{jeong:2012}. 

Experiments on material samples \cite{klein:1980,tamura:1995} showed that the assumption of local reaction is not accurate for a layer of porous material
covering a rigid base, in which case the boundary impedance evidenced angle dependence. In such case the \emph{extended reaction} assumption is preciser and improves the theoretical and experimental agreement of sound pressure level predictions \cite{franzoni:2003}. Moreover, the significance of introducing extended reaction boundaries into computer models has been demonstrated -- using phased beam tracing, the influence on the steady state sound pressure level in frequency bands was shown by Hodgson and Wareing \cite{Hodgson:2008} and on several acoustic parameters (strength, reverberation time and rapid speech transmission index) by Yousefzadeh and Hodgson \cite{yousefzadeh:2012}. Recently, Yasuda \emph{et al.} \cite{yasuda:2016} showed by wave-based modeling that the extended reaction alters the response in particular at low frequencies, which is also the frequency range of our interest. 

In the previous study \cite{prislan:2016} we have shown that in a rectangular room with a rather clean BC the lowest room modes can be correctly modeled by the RTS modeling technique. The main focus of the study was firstly to adopt the semiclassical theory and establish a connection of this elaborated concept with acoustic modeling, and secondly 
to test the behavior of the RTS technique by comparing the modeled $G$ with its analytical solution. 
Weakly damped  boundaries  were assumed in the simulation, corresponding to a real frequency- and angle-independent surface impedance (local reaction boundary) identical for all bounding surfaces. 

The aim of the present study is to validate the RTS method in a practically more relevant environment, i.e., by using more realistic BCs. Besides i) the BC used in the previous study, two additional BCs are implemented: ii) a frequency-dependent local reaction BC of a resonating membrane absorber and iii) an extended reaction BC of a porous boundary layer covering a rigid base described within the Delany-Bazley-Miki model \cite{miki:1990}. 
These particular BCs were chosen because they are well known in theoretical acoustics. On equal basis, alternative BC models can be implemented into RTS as long as the reflection coefficient can be formulated.

The results obtained by RTS are compared with FEM which is considered reliable at low frequencies. 
In a rectangular room with the same proportions as in the previous study \cite{prislan:2016}, the comparison is made for the three BCs applied homogeneously and on top of that, for their more realistic inhomogeneous distribution. The comparison includes the magnitude of the obtained pressure response and the octave band decay curves with the corresponding reverberation times. The pressure response was computed up to 300\,Hz, since at higher frequencies the FEM method would not produce correct results in acceptable time limits for the models under study.  

The article is organized as follows.  In Sec.~\ref{THEO} the BCs are introduced and in Sec.~\ref{METHO}  the most important aspects of the simulation methods are presented. In Sec.~\ref{RESULTS}, the studied cases are defined and the results are presented, compared and discussed. Main conclusions are drawn in Sec.~\ref{CONCLUSION}.

\section{Boundary conditions}\label{THEO}

\subsection{Specular reflection coefficient}\label{SPEC}

\noindent As we are concerned with the behavior of the method at lower frequencies, only specular reflections are considered. Diffuse (stochastic) reflections are not included although they are known to be significant in geometrical modeling techniques \cite{vorlander:1995,savioja:2015}. 
A link between the diffuse sound field and its geometrical interpretation was described in ref.~\cite{prislan:2014}, together with experimental tests that indicated that the complexity of the room boundaries (which produces diffuse reflections) is relevant only at higher frequencies.

In geometrical modeling techniques, the BC is implemented via the specular reflection coefficient $r$ \cite{kuttruff:2007}, which is the ratio $B/A$ of complex sound pressure amplitudes of the reflected wave, $p'({\bf r},t)=B \exp [j (\mathbf{k}'\cdot \mathbf{r}-2\pi f\,t)]$, and the incident wave,
\begin{equation}
	p({\bf r},t)=A \exp [j (\mathbf{k}\cdot \mathbf{r}-2\pi f\,t)],\label{eq:planew}
\end{equation} 
where $|{\bf k}| = |{\bf k}'| \equiv k = 2\pi f/c_0$ is the wavenumber,  $f$ is the frequency, $c_0$ the speed of sound in air 
and $j$ the imaginary unit.
Generally, $r$ is a complex quantity inducing both a magnitude as well as phase change of the reflected sound wave. 
As each sound ray is multiply reflected, its amplitude $B_i$ after the $i$-th reflection is calculated recursively as
\begin{equation}
	B_i=r\,B_{i-1},
\label{eq:B} 
\end{equation}
with $B_0=1$. 

The reflection coefficient is completely determined by an effective impedance $z$ of the boundary \cite{kuttruff:2007} and the angle of incidence $\theta$ of the ray with respect to the boundary normal,
\begin{equation}
	r(\theta)={z \cos{\theta} -z_0 \over z \cos{\theta} + z_0},
\label{eq:reflection} 
\end{equation}
where $z_0=\rho_0 c_0$ is the air impedance and $\rho_0$ the density of air ($\rho_0=1.204$ kg/m$^3$ and $c_0=343$ m/s is considered throughout this study).

Under the local reaction assumption the BC includes only the normal velocity component \cite{fahy:2001} and the effective impedance is just the specific impedance of the bounding material, which is angle-independent.
In contrast, in the case of an extended reaction boundary, its effective impedance depends on the frequency and the angle of incidence, as discussed in Sec.\,\ref{extended}. 

The reflection coefficient is directly related to the (energy) absorption coefficient $\alpha(\theta) = 1-|r(\theta)|^2$. 
The half-space solid angle average of $\alpha(\theta)$ is the diffuse sound field absorption coefficient $\alpha_{\rm diff}$, 
\begin{equation}
	\alpha_{\textrm{diff}}=2 \int_0^{\pi/2} \alpha(\theta) \cos{\theta} \sin{\theta} \,\mathrm{d}\theta,
\label{eq:paris} 
\end{equation}
which is known as the Paris formula \cite{mechel:2008}.
For common materials only $\alpha_{\textrm{diff}}$ is standardly specified (e.g., measured according to the standard ISO354 \cite{ISO354}), which opens the discussion \cite{jeong:2008} of how to determine the reflection coefficient $r$ required in phased geometrical modeling.

The computational efficiency of the RTS method is not compromised if $z$ depends on frequency and angle of incidence. In fact, in this study we put this under a direct test and introduce just these dependencies, arising naturally in the implementation of the three different BC described in detail in the following.

\subsection{Real impedance boundary}\label{real}

\noindent
The BC previously tested with RTS \cite{prislan:2016}, i.e., weakly damped boundaries with a real impedance $z_n=60\,z_0$, 
is an example of local reaction and is used also in this study for comparison. 
The corresponding reflection and absorption coefficients follow from Eqs.~(\ref{eq:reflection}) and (\ref{eq:paris}) and are frequency-independent as also shown in Fig.~1.
This BC is not particularly realistic, although rather low and frequency-independent absorption coefficients are in practice measured for massive structures with smooth finishing (e.g., painted concrete walls).
\begin{figure}[h!]
	\centering
	\includegraphics[width=0.49\textwidth]{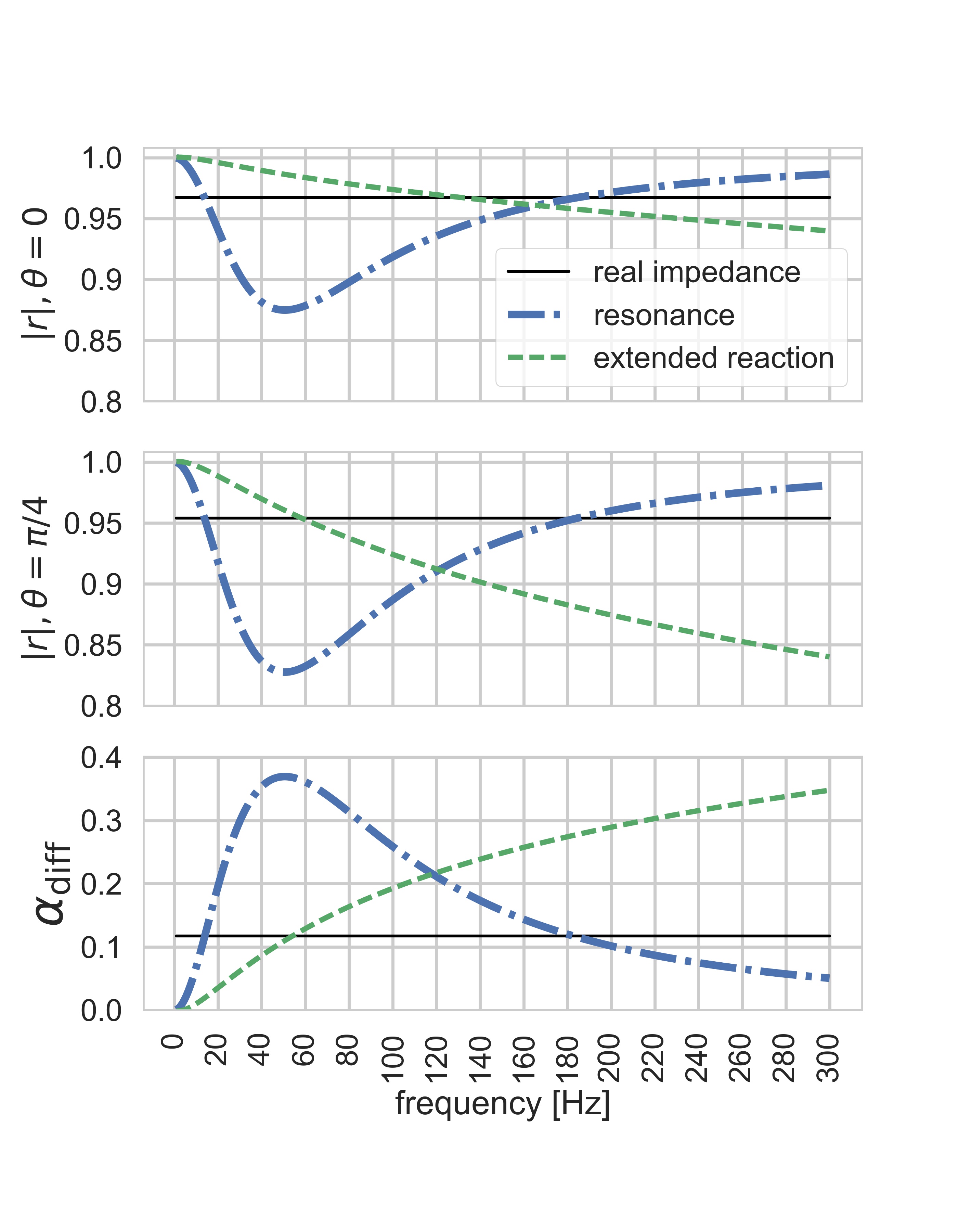}
	\caption{(Color online) 
	Absolute values of the reflection coefficients $r(\theta)$ for the real impedance, resonance and extended reaction BCs used in the simulation, for incidence angles $\theta=0$ (top) and $\theta=\pi/4$ (middle). Bottom: the corresponding diffuse sound field absorption coefficients $\alpha_{\rm diff}$ as given by Eq.~(\ref{eq:paris}).} 
	\label{fig:1}
\end{figure}

\subsection{Resonance absorber}\label{resonating}

\noindent This BC models a resonating sound absorber such as of a membrane type, characterized by its complex impedance \cite{cox:2009}
\begin{equation}
	z_{m}(f)=r_m-j\left(2 \pi f m_S-\rho_0 c_0 \cot{k d_m}\right) 
\label{eq:resonator} 
\end{equation}
consisting of specific resistance $r_m$, specific inertance $-j 2 \pi f m_S$ of the membrane mass per area $m_S$ and specific stiffness $j \rho_0 c_0 \cot{k d_m}$ of the air-filled cavity with depth $d_m$. For low frequencies ($kd_m\ll 1$), $\cot kd_m\approx 1/kd_m$ and the impedance Eq.~(\ref{eq:resonator}) is simplified to
\begin{equation}
	z_{m}(f)=r_m-j{ 2 \pi \,m_S\over f}\left(f^2 -f_m^2\right),
\label{eq:resonator2} 
\end{equation}
where 
\begin{equation}
f_m={c_0\over 2 \pi} \sqrt{{\rho_0\over m_S d_m}}
\label{eq:resonator3} 
\end{equation}
is the resonance frequency of the absorber.

The parameters used in the simulation are $m_S=10$ kg/m$^2, \,d_m = 0.14$ m and $r_m=15 \,z_0$ resulting in the resonance frequency of $f_m=50.6$ Hz. In Fig.~1, the corresponding deep in the reflection coefficient and peak in the absorption coefficient can be seen at the resonance frequency.

\subsection{Extended reaction boundary }\label{extended}
 
\noindent
A layered structure of the boundary leads to the extended reaction BC.
In our case we use a $d_p=5$\,cm thick layer of porous material covering a perfectly rigid wall.

An empirical model of rigid-frame porous materials was introduced by Delany and Bazley in 1970s \cite{delany:1970}. The model was later modified and generalized by Miki \cite{miki:1990} and is today known as the Delany-Bazley-Miki model. 
In our implementation, we follow its recent description by Yasuda \emph{et al.} \cite{yasuda:2015}.

In this model, the specific impedance of the bulk porous material $z_p$ and the wavenumber $k_p$ (dispersion relation) are given by
\begin{equation}
	z_p= z_0\, g_z \left( {f \over \sigma }\right), \quad 
	k_p=k\, g_k  \left( {f \over \sigma }\right), \label{eq:DBM1}
\end{equation}
where $\sigma$ [N$\,$s/m$^4$] is the linear specific airflow resistivity and the functions $g_z(x)$ and $g_k(x)$ are 
\begin{eqnarray}
	g_z(x)&=&1+a_zx^{q_z}-j b_zx^{q_z},  \label{eq:DBM3}\\
	g_k(x)&=&1+a_kx^{q_k}-j b_kx^{q_k},  \label{eq:DBM4}
\end{eqnarray}
with the constants $a_z=0.07, b_z=-0.107, q_z=0.632, a_k=0.109, b_k=-0.160, q_k=0.618$. 
Note that the signs of $b_z$ and $b_k$ are opposite to those of ref.~\cite{yasuda:2015}, in accord with our definition of the plane-wave time factor, Eq.~(\ref{eq:planew}).

The effective impedance of such an extended reaction boundary, formed by a rigid wall covered with a layer of the described porous material with thickness $d_p$, can be shown to be \cite{yasuda:2015}
\begin{eqnarray}
	z_e(f,\theta)&=&-j z_0 g_z ({f/\sigma}) {g_k ({f / \sigma }) \over \sqrt{g_k^2 ( {f / \sigma }) - \sin^2{\theta}}} \times \nonumber \\
	&\times& \cot{\left[ {2\pi f d_p\over c_0} \sqrt{g_k^2 ( {f / \sigma }) - \sin^2{\theta}} \right]}\label{eq:DBM5}
\end{eqnarray}
and depends on the frequency $f$ as well as on the angle of incidence $\theta$.
To complete the picture, $\sqrt{g_k^2  - \sin^2{\theta}}/g_k = \cos\theta'$, where $\theta'$ is the (complex) angle of refraction in the porous layer.

The value $\sigma=500$\,Ns/m$^4$ was used in the simulation and the corresponding reflection coefficient Eq.~(\ref{eq:reflection}) and diffuse field absorption coefficient Eq.~(\ref{eq:paris}) are plotted in Fig.~1. It can be seen that in the considered frequency range the reflection coefficient monotonically decreases while the absorption coefficient monotonically increases with frequency.

\section{Methods}\label{METHO}

\subsection{Pressure response function}\label{PRESSURE}

\noindent 
The Green's function $G({\bf r},{\bf r}_0,k)$ (the response to a dimensionless point source of strength 1 located at ${\bf r}_0$) of the Helmholtz equation \cite{prislan:2016} satisfies
\begin{equation}
	-\left(\nabla^2 + k^2\right) G({\bf r},{\bf r}_0,k) = \delta({\bf r}-{\bf r}_0),
	\label{eq:greenpdef0}
\end{equation}
while the physical \emph{sound pressure response} function $G_p({\bf r},{\bf r}_0,k)$, i.e., the pressure response to a pointlike sound source of strength $(\dot{\phi}_m)_0$, satisfies 
\begin{equation}
	-\left(\nabla^2 + k^2\right) G_p({\bf r},{\bf r}_0,k) = (\dot{\phi}_m)_0\,\delta({\bf r}-{\bf r}_0).
	\label{eq:greenpdef}
\end{equation}
The amplitude  $(\dot{\phi}_m)_0$ of the rate of change of the radiative mass flux, i.e., the radiative mass acceleration generated by the sound source, is directly connected with the acoustic power $P$ emitted by the source,
\begin{equation}
	(\dot{\phi}_m)_0 = \sqrt{8\pi\rho_0 c_0 P}.
	\label{eq:pressure_source}
\end{equation}
This relates the Green's function $G$, which is furnished by RTS, with the sound pressure response function $G_p$:
\begin{equation}
	G_p({\bf r},{\bf r}_0,k) = \sqrt{8\pi\rho_0 c_0 P}\, G({\bf r},{\bf r}_0,k). \label{eq:GFp}
\end{equation}

\subsection{RTS method}

\noindent
The RTS method resides on the semiclassical approximation of quantum mechanics and in our previous publication \cite{prislan:2016} we have demonstrated that this concept can be transferred also to acoustics. 
The resulting acoustic RTS method is a frequency domain method, the main advantage of which is that it yields directly the Helmholtz equation Green's function $G$ of the room.

From the technical perspective it is a ray-tracing method, where sound rays are propagated from the source ${\bf r}_0$ in random directions, upon hitting a boundary they are reflected according to reflection rules, and are in the end summed up in a point (region) of interest. In our case, the rays are emitted in random directions with isotropic probability distribution. At the boundary they are specularly reflected and the pressure amplitude factor changes as defined by Eq.~(\ref{eq:B}), with the reflection coefficient $r$ corresponding to the BC. 

The segments of a multiply reflected ray constitute a \emph{trajectory}. If a trajectory crosses the observation region centered at $\mathbf{r}$, it qualifies as a \emph{path}, i.e., a valid connection between the source and the observation region. The Green's function $G$ is numerically constructed as \cite{prislan:2016}
\begin{equation}
	G(\mathbf{r},\mathbf{r}_0,k)={1\over \pi R^2 N} \sum_l^L B_{l,i} \,d_l\,  \exp{\left[j\, k \,d_l\right]}, \label{eq:numgreen}
\end{equation}
where
\begin{itemize}
	\item $N$ is the total number of sound rays emitted from the point souce at ${\bf r}_0$,
	\item $R$ is the radius of the spherical observation region centered at $\mathbf{r}$, 
	\item $L$ is the total number of paths $l$ with lengths $d_l$, that are identified by the simulation,
	\item $B_{l,i}$ is the complex amplitude of the sound ray of path $l$ after $i$-th reflection. 
\end{itemize}
The factor $d_l$ represents the density of paths for the three-dimensional case of flat or piecewise flat boundaries, where (de)focusing effects are absent \cite{prislan:2016}.

\subsection{FEM method}\label{FEM}

\noindent 
The same room and BCs were also modeled using the PDE FEM solver COMSOL. There we used a monopole source with acoustic power of 1\,W to obtain the complex pressure response $G_p$ to a source with unit acoustic power. 
Taking into account the connection Eq.~(\ref{eq:GFp}), $G_p$ and the response $G$ obtained by RTS can be directly compared.

\section{Results and discussion}\label{RESULTS} 

\noindent 
The dimensions of the room were $a\times b \times c=4.215\,\textrm{m} \times 3.647 \,\textrm{m} \times 3\, \textrm{m} $ as in our previous study \citep{prislan:2016}. 
To exclude that any conclusions would be based on the particularity of a chosen position,
the simulation was performed for two pairs of source/observation points, Fig.~2, designated as \textit{corner} and \textit{inner} positions. 

Rather than fixing the number of reflections of each ray, we introduce a low-amplitude cutoff: a ray sequence is terminated when $|B_i|< 10^{-6}$ independent of the frequency. For the chosen geometry and absorption strengths this gives typically between 500 and 900 reflections per sound ray.

The size of the observation region was $R=0.2$\,m and the number of emitted sound rays was $N=6\,758\,400$. 
Without parallelization, this simulation would be running for $\sim$ a month on a single core processor of a desktop computer (no code optimization). In our case the code was parallelized and was running on a cluster for 4 days.
We however emphasize that practically valuable results are obtained already with drastically less rays and reflections, reducing the computational resources to a fraction.

Four BCs were tested: real impedance (Sec.\,\ref{real}), resonance (Sec.\,\ref{resonating}), extended reaction (Sec.\,\ref{extended}) and mixed. In the first three cases the same BC is applied to all boundaries (homogeneous BC), whereas in the mixed case the three BC are distributed as shown in Fig.~2.

Adjusting the rigid wall position of the extended reaction boundary such as to compensate for the 5\,cm layer of porous material,
the inner room dimensions were the same for all studied cases. In this way the rays were reflected from the same planes in all cases and only one RTS simulation was needed, in which the four studied BCs were computed simultaneously. 
To the contrary, for each BC a separate FEM simulations had to be performed in COMSOL. 

The pressure response $G_p$ was calculated with both methods at the same frequencies in 0.05\,Hz steps up to 300\,Hz. The FEM meshing was performed automatically with the maximum allowed mesh spacing set to 0.34\,m. This limit granted at least 5 mesh points per wavelength up to 200\,Hz. An additional verificatory simulation with a much finer mesh did not deliver any observable changes.

\begin{figure}[h!]
	\centering
	\includegraphics[width=0.49\textwidth]{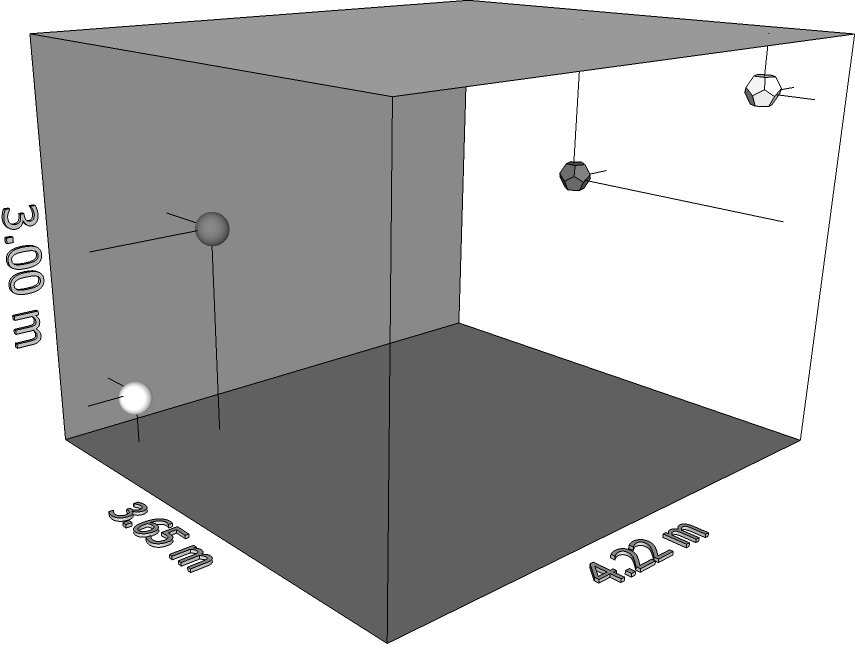}
	\caption{A sketch of the room with sound sources (dodecahedrons) and observation points (spheres) for the two studied source/observation pairs. 
The corner positions (white) are situated in diagonally opposite corners 0.402\,m away from each wall, while the inner positions (gray) are at $\mathbf{r_0}=(3.6, 1.8, 1.1)$\,m and $\mathbf{r}=(1, 0.6, 1.6)$\,m. 
In case of the mixed BC, the floor (dark gray) is imposing the real impedance BC, the ceiling and one of the walls (light gray) the extended reaction BC, while the remaining walls (transparent) are imposing the resonance BC.} 	\label{fig:2}
\end{figure}

\subsection{Pressure response}\label{RESULTS:GF}

\noindent
In Fig.~3, the magnitudes of the pressure responses obtained with RTS and FEM for the four BCs and both positions are presented in dB scale up to 300\,Hz. We observe the following.
\begin{figure}[h!]
	\centering \includegraphics[width=0.48\textwidth]{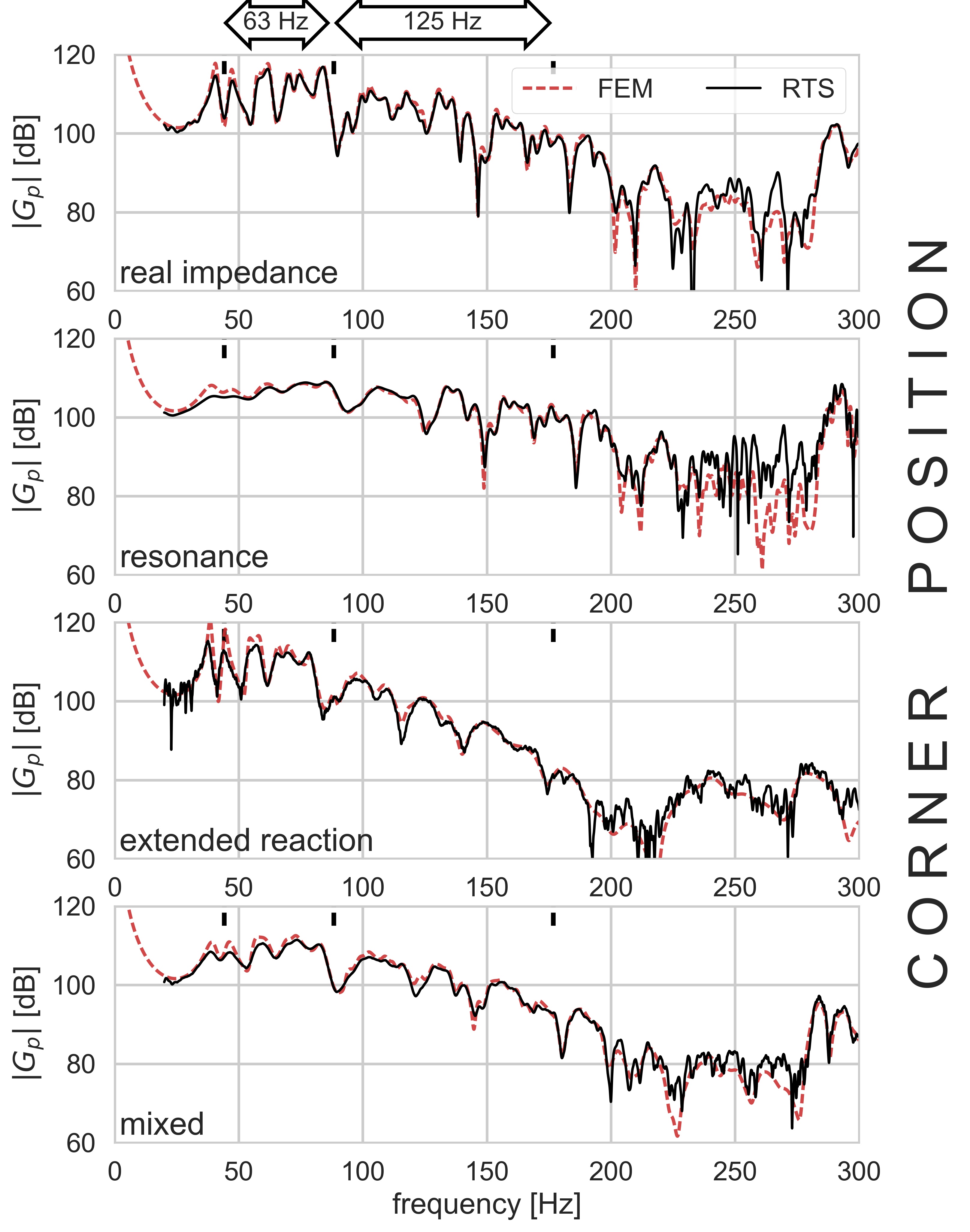} \newline
	\centering \includegraphics[width=0.48\textwidth]{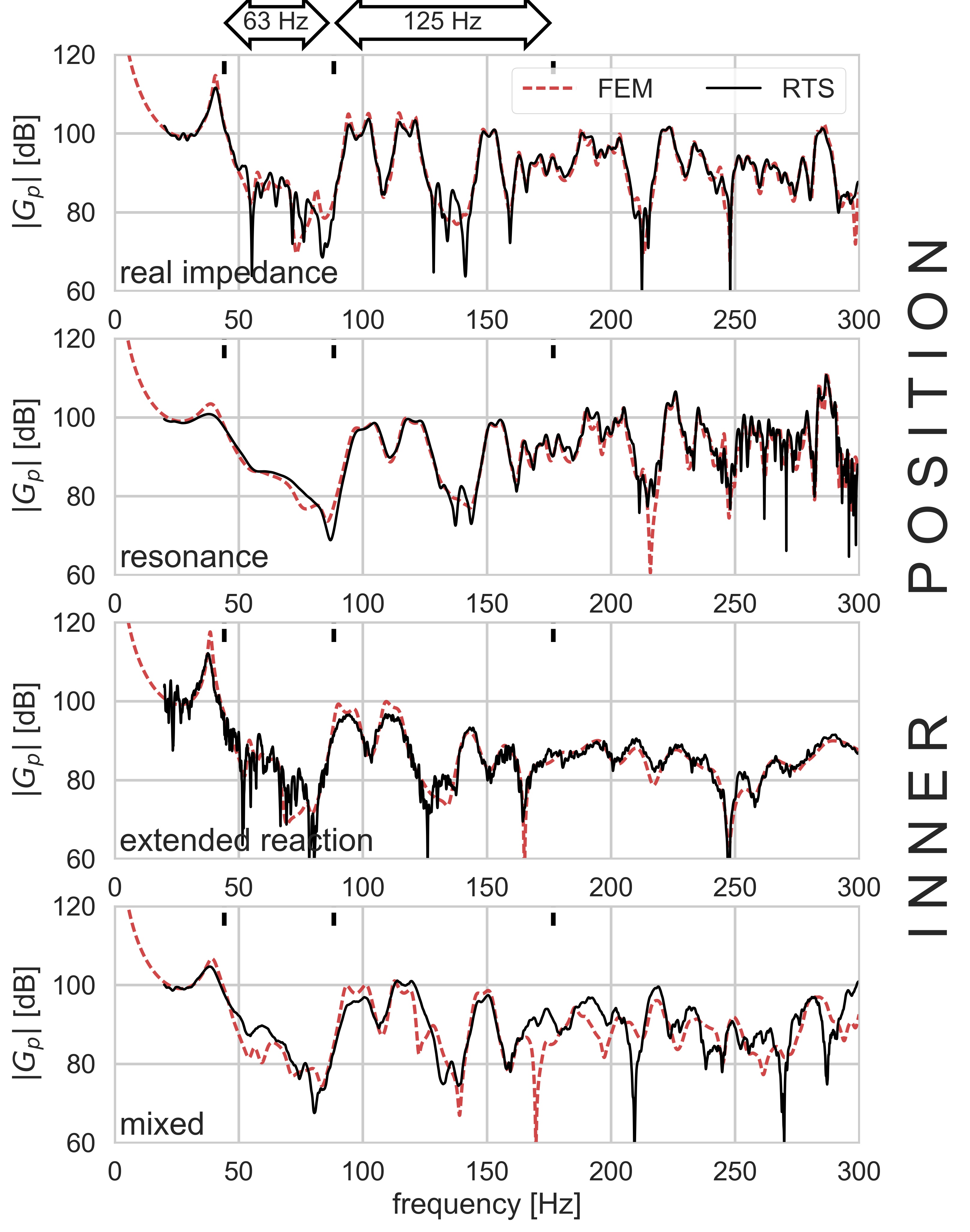}
	\caption{(Color online) The absolute value $|G_p|$ of the pressure response modeled by RTS (solid black) and FEM (dashed red) for the corner (top four graphs) and inner (bottom four graphs) positions and the four BCs as indicated.
	Horizontal arrows outline two octave bands for which the reverberation times are determined.} 	
	\label{fig:3}
\end{figure}
\begin{itemize}
\item For all studied cases the normalization of $G_p$ is correct. In most cases also the peaks and deeps of $|G_p|$ coincide well in frequency, level and width. 

\item A general exception is present for the lowest room modes (the peaks  below 70\,Hz) which are systematically slightly underestimated by RTS. This phenomenon was observed and discussed already in the previous study \cite{prislan:2016} and seems to be an inherent feature of the semiclassical approximation.

\item A noisy behavior of RTS can be observed in frequency zones where the magnitude of $G_p$ is lower. This might be an indication of a noise floor that is not of statistical nature, but rather of numerical origin, i.e.,  it is not readily lowered by using more rays.
A possible source of such numeric noise could be the finite floating point accuracy of determining the reflection angles and coordinates of boundary impacts. Although relatively small for double precision, the error increases with every reflection and propagation and is amplified by two or three orders of magnitude over the whole paths. Since the paths are long but at the same time confined to the room, the rather small relative error of a long path can translate into a substantial error relative to the size of the room.
 
\item Due to the higher absorption of the extended reaction boundaries above 120\,Hz (see Fig.~1), $|G_p|$ is lower in this frequency range in comparison with the other BCs. Therefore, again, more noise in $G_p$ computed by RTS can be observed, which is especially evident for the inner position. 
Moreover, for this BC the curves appear more erratic even for $|G_p|$ as high as $\sim 100$\,dB, which is not observed for the other BCs. 
As the extended reaction BC is more involved, i.e., it requires more algebraic operations and function evaluations to determine the reflection coefficient of each reflection, this supports the finite floating point accuracy as a candidate responsible for this error.

\item Moreover, a noisy behavior of RTS for the extended reaction BC is observed also in the frequency range below 40\,Hz, most probably owing to the lack of absorption there ($\alpha_{\textrm{diff}}<0.05$, Fig.~1). As this noise is present only for very low absorption, it is not perceived as a practical limitation of the RTS method. 

Further investigations where we modeled more reflections have revealed that this noise is reduced only if also more rays are emitted at the same time.
To explain this observation, we should bear in mind that rays diverge during propagation and hence their population density decreases. For the noise to average out, it is required that portions of the phase space \citep{prislan:2016} that significantly contribute to $G$ are adequately represented. Therefore, in case of low absorption when the amplitude of rays remains higher even after many reflections and thus still significantly contributes to $G$, more rays must be emitted to keep their population density adequate also in these distant parts of the paths.

\item For the mixed BC, in the corner position the agreement between both methods is equally good as for the other BCs and no peculiarities are observed. In contrast, in the inner position deviations are severer. In particular the deeps between 160\,Hz and 250\,Hz do not coincide. We find this phenomenon interesting, as it is not systematic, i.e., it occurs only in one position and in a limited frequency range. 
Performing a FEM simulation on a finer mesh on one hand, and testing the influence of RTS parameters (number of reflections, number of emitted sound rays, radius of the observation region) on the other did not bring any evident improvement.
At this stage we conclude that this peculiarity should be systematically addressed in further more specific studies -- by testing more positions and more variety in the distribution of different BCs.

\end{itemize} 

In case of the extended reaction BC the peaks occur at lower frequencies in comparison to the other two BCs. 
This frequency shift results from the fact that i) for the extended reaction BC the linear size of the room without the inside porous layer is $\sim 3$\,\% larger and ii) the sound speed in the porous layer is lower ($\sim 1$\,\% at 10\,Hz up to $\sim 8$\,\% at 300\,Hz). Both effects add up in lowering the resonance frequencies.
The frequency shift is the smallest for the lowest modes and increases for higher modes. It is most evident in the central position for the peaks in the 90-170\,Hz range. In the mixed BC case, the shift is less visible.

\subsection{Decay curves and reverberation time}\label{RESULTS:EDC}

\noindent
The impulse response was calculated as an inverse Fourier transform of the octave-band-filtered (class 1 filters in agreement with EN 61260-1:2014 \cite{EN61260}) pressure response. The two octave bands considered are indicated in Fig.~3.
Next, the 0.8\,s window of the squared impulse response was reverse integrated \cite{schroeder:1965} and the decay curve was obtained, Fig.~4. 
By a linear fit to the -5\,dB to -25\,dB part of the decay curve, $T_{20}$ was calculated as defined by the standard ISO3382 \cite{ISO3382}, serving as an estimate for the reverberation time.
\begin{figure}[h!]
	\centering 	\includegraphics[width=0.48\textwidth]{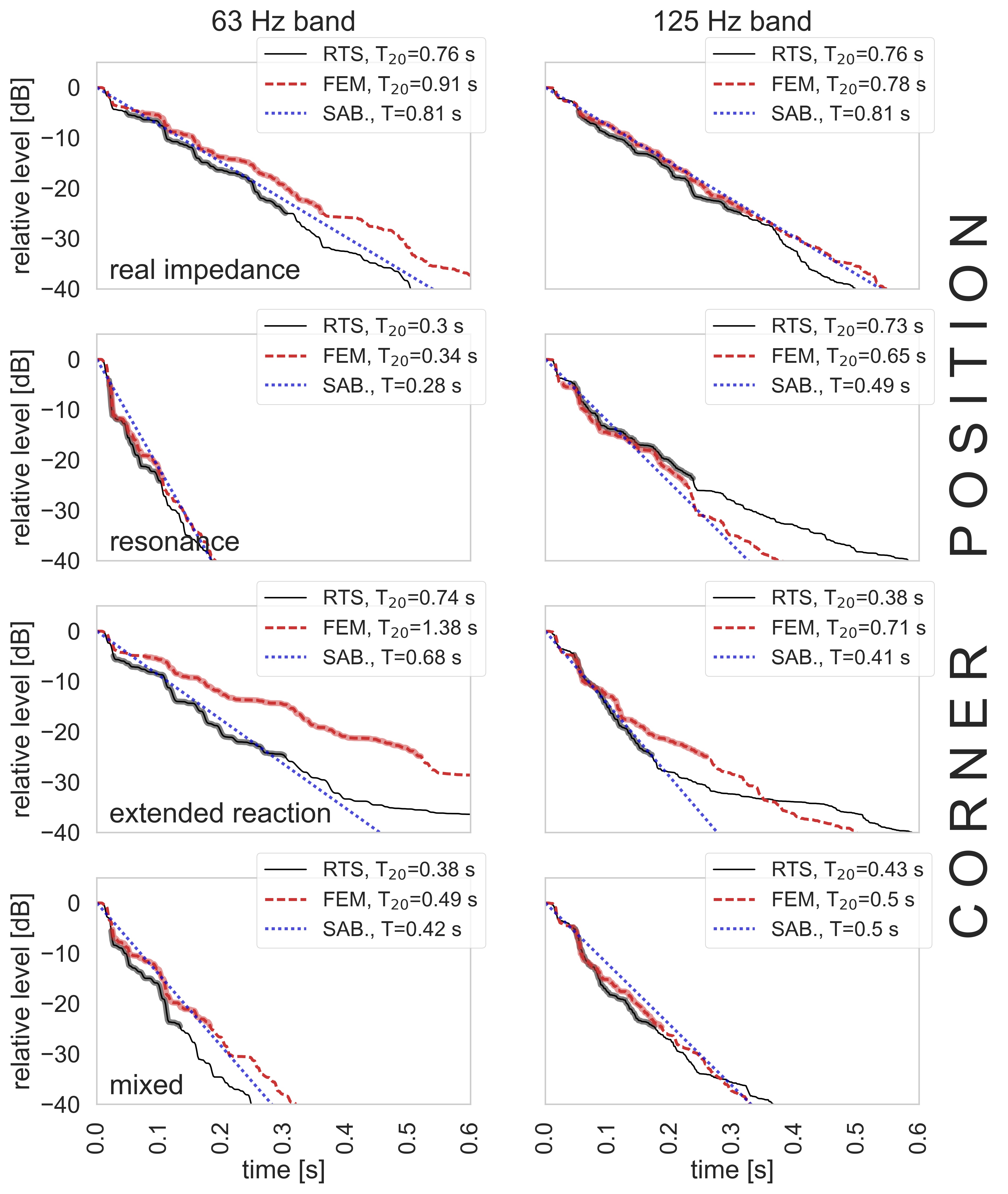}\newline
	\includegraphics[width=0.48\textwidth]{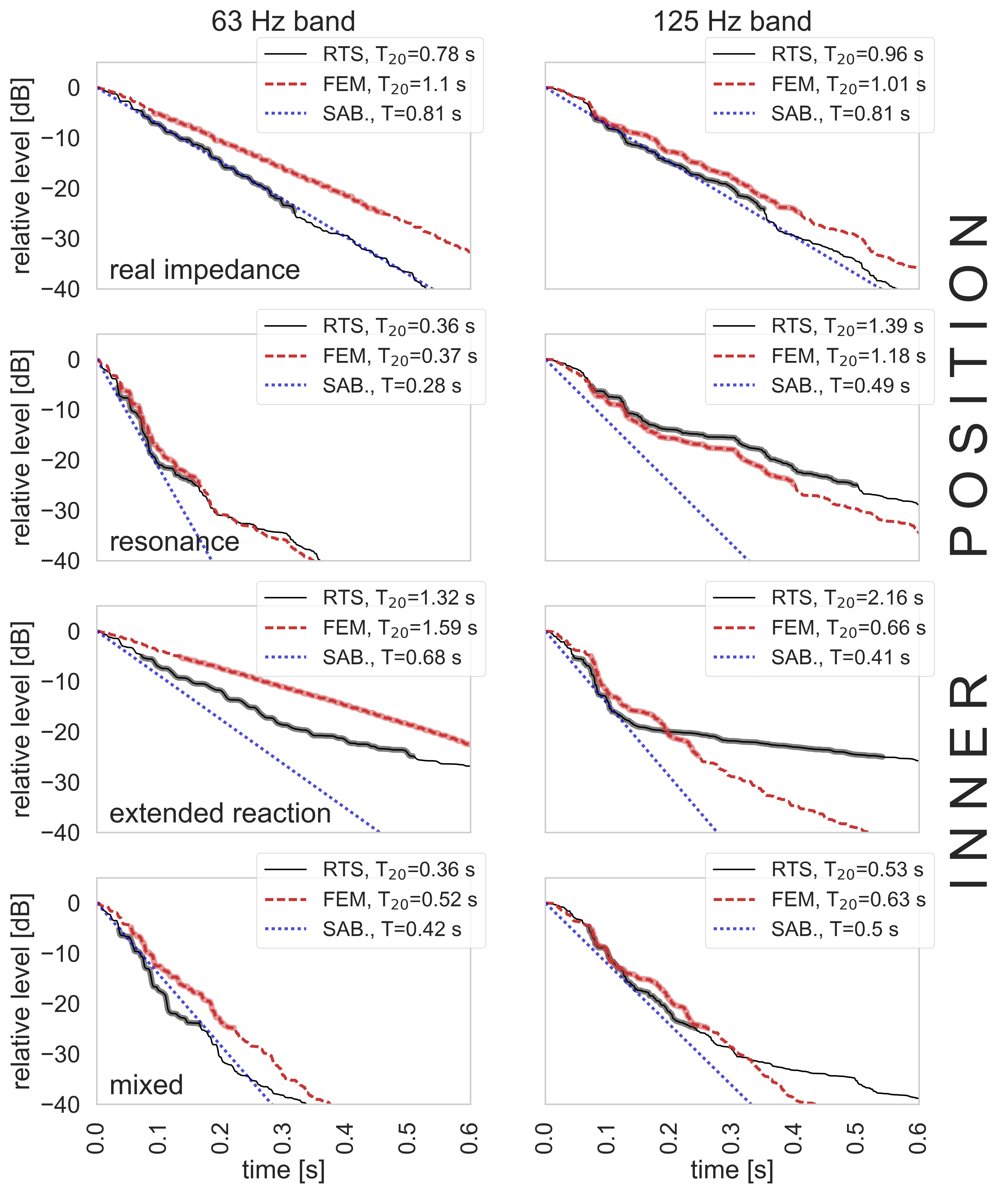}
	\caption{(Color online) RTS (solid black), FEM (dashed red) and Sabine (SAB., dotted blue) sound pressure decay curves and the corresponding reverberation times in the two octave bands for the four BCs, computed in the corner (top) and inner (bottom) positions. Bold sections of the numerical curves correspond to the -5\,dB to -25\,dB ranges used to extract T$_{20}$ by a linear fit.}
	\label{fig:4}
\end{figure}

For comparison the reverberation times were calculated also by the Sabine equation using absorption coefficients as octave band averages of $\alpha_{\textrm{diff}}$ given by the Paris formula Eq.~(\ref{eq:paris}). The so obtained average absorption coefficients $\bar{\alpha}_{\rm diff}$ for each BC are listed in Table~\ref{tab:a}.

\begin{table}[h!]
\begin{tabular}{c|c|c}
& \multicolumn{2}{c}{$\bar{\alpha}_\textrm{diff}$}\\  
~\parbox{1.4cm}{BC} &
~\parbox{1.4cm}{63 Hz} &
~\parbox{1.4cm}{125 Hz}  \\
\hline
real impedance & $0.12$ & $0.12$ \\
resonance & $0.27$ & $0.15$  \\
extended reaction & $0.13$ & $0.24$  \\
\end{tabular}
\caption{The octave-band-averaged diffuse absorption coefficients for the three BCs. }
\label{tab:a}
\end{table}

The decay curves and reverberation times for both simulation points are presented in Fig.~4. The following can be observed.
\begin{itemize}

\item The obtained dynamic range is adequate to determine T$_{20}$ everywhere with the exception of the RTS extended reaction BC inner position in the 125\,Hz band. This case corresponds to the region of low $|G_p|$, where the highest amount of noise is present, Fig.~3. As a consequence, the reverberation time determined by RTS is highly overestimated and is thus irrelevant. 

The largest discrepancy between the RTS and FEM $T_{20}$, $17\,\%$-$227\,\%$, is found for the case of the extended reaction BC. Again it is due to the noisy RTS response, Fig.~3, which restrains the dynamic range of the decay curves. In contrast, for the other BCs the average (maximum) discrepancy between the RTS and FEM reverberation times is $15\,\%$ ($31\,\%$).  Particularly, for the 125\,Hz band the average discrepancy is only $11\,\%$.

\item The decay curves are most linear for the real impedance BC in the 125\,Hz band. In these cases the difference in $T_{20}$ computed by FEM and RTS is only $3\,\%$ and $5\,\%$.

\item Characteristically, for the 63\,Hz band the $T_{20}$ computed by RTS is shorter in comparison with FEM for all studied cases. This is due to the underestimation of the $|G_p|$ peaks by RTS, Fig.~3. In fact, lowered peaks correspond to an increased relative width of the modes, which imitates an increased absorption and thus a shorter reverberation time.

\item As expected, the Sabine reverberation time, which is by definition equal for both positions, is closer to $T_{20}$ determined by RTS and FEM for the corner position, where there are more peaks in the pressure response, Fig.~3, and the modal overlap is higher. At a higher overlap the diffuse sound field assumption -- the basic assumption behind the Sabine equation -- is better justified. 

\end{itemize}

\noindent
It thus follows 
that the RTS method is suitable for the determination of the reverberation time, provided the dynamic range of the decay curve is adequate.
If the frequency band includes the lowest room modes, the RTS reverberation time can be underestimated by up to 30\,\% in absolute. Its relative behavior depending on the positions of the source and the listener nevertheless remains significant also in this case.

\section{Conclusion}\label{CONCLUSION}

\noindent
We have performed a stark test of the acoustic RTS (ray-tracing semiclassical) method for a wide selection of realistic boundary conditions, confronting it face-to-face with the results of a full wave FEM method which is considered accurate in the studied lower frequency range.

First of all, we have shown that the pressure responses obtained by RTS and FEM agree in their normalization, i.e., there is no scaling uncertainty in the RTS-computed response.
Moreover and most importantly, for the majority of the studied cases the peaks and deeps of the spatially dependent responses coincide remarkably well in frequency, level and shape. The agreement is systematic for all studied BCs with the exception of the mixed BC in the inner position, where in a limited frequency range the deeps in the response do not coincide.

Some further particularities of the RTS method were exposed -- the underestimation of the peaks corresponding to the lowest modes and the presence of noise in case of low $|G_p|$ values as well as for very reflective boundaries. 
Generally we can conclude, that for the case of the rectangular room and tested BCs, RTS is an adequate method for modeling the spatially dependent pressure response.

Excluding the case of the pure extended reaction BC, the average discrepancy between FEM- and RTS-computed reverberation times was $15\,\%$. Owing to the weak absorption at low frequencies, the extended reaction BC produced a strong noise in this part of the pressure response, which restrained the dynamic range of the decay curves. When the extended reaction BC was combined with other BC types that supplied enough absorption at low frequencies, this noise problem disappeared.

Best agreement of the estimated $T_{20}$ was obtained for the real impedance BC in the 125\,Hz band. In this case the FEM- and RTS-computed reverberation times differed only by 3\,\% and 5\,\% for both simulation positions, respectively. Overall, the study showed that RTS is an adequate method for predicting the reverberation time in rectangular rooms at low frequencies. At the same time, the importance of the adequate dynamic range of the computed decay curves was exposed.

In conclusion, our study validates the RTS method for a set of practically relevant boundary conditions, thereby further promoting it for low-frequency geometrical acoustic modeling in rectangular rooms. 
Some uncertainties in the case of inhomogeneous BCs were revealed, which should be systematically addressed in the future. 
Future studies should also validate and characterize the RTS method for general room geometries, including in particular (de-)focusing effects of curved boundaries.

\section*{Acknowledgments}

\noindent
The Authors acknowledge partial support of the Slovenian Research Agency (Grant J1-7435), the Slovenian Ministry of Higher Education, Science and Technology and the European Regional Development Fund.

%
%
%
%
			
\end{document}